\documentclass[sigconf]{acmart}

\usepackage{tikz}
\usepackage{amsmath}

\usepackage{filecontents}
\usepackage{algorithm,algpseudocode,ifmtarg}
\usepackage{acronym}
\usepackage{listings}
\usepackage{todonotes}
\usepackage{graphicx}

\makeatletter
\renewcommand{\ALG@name}{Code}
\newcommand{\atoprule}{\hrule height.8pt depth0pt \kern2pt} 
\newcommand{\amidrule}{\kern2pt\hrule\kern2pt} 
\newcommand{\abottomrule}{\kern2pt\hrule\relax}
\newcommand{\algcaption}[2][]{%
	\refstepcounter{algorithm}%
	\@ifmtarg{#1}
	{\addcontentsline{loa}{figure}{\protect\numberline{\thealgorithm}{\ignorespaces #2}}}
	{\addcontentsline{loa}{figure}{\protect\numberline{\thealgorithm}{\ignorespaces #1}}}%
	\atoprule
	\textbf{\fname@algorithm~\thealgorithm}\ #2\par 
	\amidrule
}
\makeatother

\acrodef{CFG}{Control-Flow Graph}
\acrodef{DSL}{Domain Specific Language}
\acrodef{ILP}{Integer Linear Program}
\acrodef{PCFL}{Partial Control-Flow Linearization}
\acrodef{RTL}{Register Transfer Language}
\acrodef{WCET}{Worst-Case Execution Time}

\newcommand{\astree}{Astrée}
\newcommand{\timingprofiler}{TimingProfiler}
\newcommand{\theproject}{FreeSBee}

\acmConference[HASP 2026]{Hardware and Architectural Support for Security and Privacy}
\acmYear{2026}
\copyrightyear{2026}

\begin{document}

\date{}

\title{\Large \bf \theproject{}: A Constant-Time Compilation Pipeline}

\author{Julian Ganz}
\author{David Knothe}
\author{Sebastian Reiter}
\affiliation{%
  \institution{FZI Research Center for Information Technology}
  \city{Karlsruhe}
  \country{Germany}
}

\author{David Oswald}
\author{Thorben Schneider}
\affiliation{%
  \institution{Kasper $\&$ Oswald GmbH}
  \city{Bochum}
  \country{Germany}
}

\author{Simon Wegener}
\affiliation{%
  \institution{AbsInt Angewandte Informatik GmbH}
  \city{Saarbrücken}
  \country{Germany}
}

\lstset{
    language=C,
    basicstyle=\rmfamily,       
    columns=fullflexible,       
    keywordstyle=\bfseries,     
    commentstyle=\itshape,      
    keepspaces=true,            
    tabsize=4,                  
    showstringspaces=false,     
    frame=lines,                
    captionpos=t,
    alsoletter={_0123456789}, 
    otherkeywords={uint8_t, __attribute}
}

\begin{abstract}
Constant-time execution is an important security property, especially for IoT devices, that prevents the leakage of sensitive information over timing side-channels.
However, achieving constant-time execution is very hard when going down to the cycle level, at least on realistic devices.
A lot of research is concerned with producing tools that either determine whether code is constant-time
or that ensure it is compiled into a constant-time binary, with various degrees of accuracy.
In this work, we present and evaluate a full workflow for hardening against timing side-channels during compilation.
It starts with taint analysis to determine the code locations that depend on sensitive information.
These are then the object of automated control-flow transformations inside the compiler to seal possible timing side-channels.
We evaluate our workflow in detail on a realistic application running on a RISC-V core
using both static timing analysis and statistical methods.
Even though we do not reach cycle-accurate constant-time execution,
our statistical results suggest that our hardened binary is safe against timing attacks.
\end{abstract}

\maketitle

\section{Introduction}
\label{sec:intro}

Due to their widespread areas of application, ranging from automotive and
medical devices to resource-constrained deployments in smart homes and
industrial automation, modern connected embedded systems such as IoT devices
must implement strong functional security guarantees.
However, even systems with robust security mechanisms in place may leak sensitive information through timing side-channels.
These leaks are caused by variances in a program's execution time that depends on confidential data \cite{kocher1996side-channels}.
Therefore, it is essential to ensure that the execution time of a sequence of instructions is independent of sensitive data, such as cryptographic keys.

Software with this property is considered to be ``constant-time'' with respect to the secrets. 
Traditional approaches at source code level aim to achieve constant-time behaviour by using
well-defined patterns and code guidelines, sometimes with the intention of
preventing compiler optimizations that would introduce variances in execution time. 
However, such approaches often lead to less expressive code that also depends on specific compilers and
compilation settings,
making them increasingly impractical in real-world applications \cite{Pornin2025Pessimist}.

Modern approaches encompass dedicated validators, compiler passes, or even
\acp{DSL} to enforce constant-time properties after compilation.
For example, compilers could be adapted to recognize and respect annotations that
express a constant-time requirement with respect to a secret.
Such a compiler would avoid optimizations or transformations that
introduce execution time variability, but only at the annotated places,
and could even eliminate some timing fluctuations that still exist in the code as-is.

In this paper, we present an implementation and evaluation of a modern
compiler-based approach to hardening software against timing side-channels.
Our approach combines an information-flow analysis with a code transformation pass in the \mbox{CompCert C Compiler}
to automatically eliminate execution time fluctuations from a program's sensitive execution paths.
The compiler transformation pass that enforces constant-time implementation is validated on a test application using AES crypto functions,
employing (a) statistical measurements and (b) static timing analysis at the binary level.

In this work, we mainly consider execution time variability introduced by control
flow and, to some extent, caching.
We acknowledge possible operand dependent variability in the execution time of
individual instructions aside from control-transfer and load/store instructions.
However, this work focuses on a workflow that can be deployed without further
assumptions or knowledge of the target hardware.

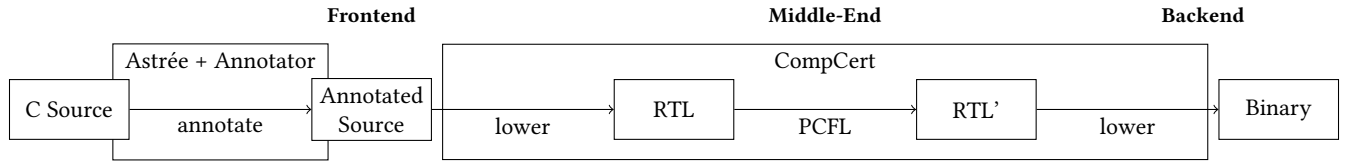
\begin{figure*}
    \center
    \begin{tikzpicture}[artifact/.style={rectangle,draw,minimum width=5em,minimum height=2.5em,align=center,fill=white}, stage/.style={font=\small\bfseries}]
        \node at (10,0.1) [rectangle,draw,minimum width=32em,minimum height=4em,text depth=3.5em,behind path] {CompCert};
        \node at (2,0.1) [rectangle,draw,minimum width=9em,minimum height=4em,text depth=3.5em,behind path] {\astree{} + Annotator};

        \node (src) at (0,0) [artifact] {C Source};
        \node (ann) at (4,0) [artifact] {Annotated\\Source};
        \node (rtl1) at (8,0) [artifact] {RTL};
        \node (rtl2) at (12,0) [artifact] {RTL'};
        \node (bin) at (16,0) [artifact] {Binary};

        \draw [->] (src.east) -- (ann.west) node [midway,below,draw=none] {annotate};
        \draw [->] (ann.east) -- (rtl1.west) node [midway,below,draw=none] {lower};
        \draw [->] (rtl1.east) -- (rtl2.west) node [midway,below,draw=none] {PCFL};
        \draw [->] (rtl2.east) -- (bin.west) node [midway,below,draw=none] {lower};

        \node at (4,1.25) [stage] {Frontend};
        \node at (10,1.25) [stage] {Middle-End};
        \node at (15,1.25) [stage] {Backend};
    \end{tikzpicture}
    \caption{Pipeline to eliminate timing side-channels using information-flow analysis and tailored compilation.}
    \label{fig:pipeline}
\end{figure*}

\section{Related Work}
\label{sec:related-work}

Pornin \cite{Pornin2025Pessimist} elaborates on why manual techniques for
writing constant-time code have become unfeasible.
They describe how optimizations performed by existing compilers may introduce
side-channels, and how techniques attempting to prevent such optimizations fail
as compilers perform deeper and deeper program analysis.
They also show how existing CPU hardware may introduce side-channels during
execution.

Several tools exist that determine whether or not a program leaks information
via timing side-channels, both on the source code level and the binary level
\cite{dude-ct, almedia2016ct-verif}.
While the assessments on the source code level are only useful under strong
assumptions regarding the compiler, assessments on the binary level eliminate
the threat of additional undetected leaks introduced during compilation.

Welch's \emph{t}-test is a statistical tool that is commonly used
for side-channel analysis \cite{schneider22}.
In our evaluation, we use both Welch's \emph{t}-test on real measurements
as well as a static timing analyzer,
comparing best-case and worst-case execution time.
Using the static timing analyzer we can determine whether the binary is
constant-time according to the used timing model,
and using Welch's \emph{t}-test we can empirically show that
any remaining timing variation is unrelated to any secrets.

There exist a few tools and compiler passes that try to remove timing leaks,
like Constantine
\cite{borrello2021constantine}, SC-Eliminator \cite{wu2018SC-Eliminator}, or Lif
\cite{soares2021lif}.
These tools try to respect memory safety and use different algorithms for
if-conversion.
Soares et~al.~\cite{soares2023PCFL} presented an improved version of Lif based
on \ac{PCFL} \cite{MollHack2018PCFL}, a global and
efficient if-conversion algorithm that can handle arbitrary loops.
There also exist compilers that provably ensure the \emph{preservation} of the constant-time property \cite{barthe2020compcert-ct-preserving}.

Our work combines several of these aspects:
we present a full pipeline ranging from taint analysis to compilation that allows
the user to control exactly where \ac{PCFL} is performed.
Using static analysis and dynamic tests, we showcase the effectiveness of our approach on a realistic application running on a RISC-V core.
Additionally, part of our implementation has been formally verified, as presented in a separate paper \cite{knothe2025-pcfl}.

\section{Workflow}
\label{sec:workflow}

Our transformation pipeline for eliminating timing side-channels in C source code is split into three phases, which are shown in Figure~\ref{fig:pipeline}.
Initially, the user manually marks secret values and variables in the original source file using a custom C attribute called \verb+secret+.

The front-end---realised by the tools \astree{} and Annotator---processes the annotations and performs an information-flow analysis to propagate tainted values throughout the entire code base.
The CompCert compiler acts as middle- and backend.
A compiler pass in the middle-end finds and eliminates timing side-channels in the \ac{CFG} via linearization.
In the backend, the transformed intermediate representation is lowered to the target architecture, which is RISC-V in our example.

\subsection{Information-Flow Analysis}

Taint analysis, or information-flow tracking, is a well-known code analysis technique that is mainly used to detect software security vulnerabilities.
It tracks the flow of data from given sources, like unsanitized user input or sensitive information like encryption keys,
and alerts when such data enters a specific sink.
While dynamic taint analysis tags sensitive data in memory during the execution of the program and observes how the data is accessed,
static taint analysis considers the program’s \ac{CFG} and collects the set of values that could possibly depend on the tainted sources.

\astree{} \cite{BlanchetCCFMMMR03} is a static source code analyzer based on abstract interpretation.
It soundly over-approximates value ranges and tracks all accesses to variables.
\astree{} includes a dedicated abstract domain for taint analysis \cite{KaestnerMWMF22}, which we use to identify variables that depend on values that have been marked by the user via the \verb+secret+ attribute.
The analysis yields a structured report that documents taint propagation and dependencies, which is then processed by the Annotator tool, adding additional \verb+maybe_tainted+ attributes to all possibly tainted function parameters and global variables throughout the source code.


\subsection{Compilation and Hardening}

CompCert \cite{leroy2009compcert} is a formally verified optimizing C compiler, which we modified in order to analyze and process the taint attributes at the \ac{RTL} level,
which is represents the code as a \ac{CFG} operating on infinitely many abstract registers.
There, a hardening algorithm is applied.
Because the taint analysis marks all affected function parameters, it suffices to harden functions independently of each other.

After the constant-time hardening, the program is lowered further, eventually resulting
in a binary.
The constant-time property is preserved by these lowering passes
\cite{barthe2020compcert-ct-preserving}.

For hardening, we implemented \acf{PCFL} \cite{MollHack2018PCFL} \cite{soares2023PCFL}, a global if-conversion algorithm that transforms a \ac{CFG} into one
that does not branch on secret conditions, thereby redirecting edges where required and
trying to linearize as little additional branches as possible.
For example, a simple \textit{if}-condition that depends on a secret will be converted into linear code
where both branches are \textit{predicated} in such a way that the code behaves identical as before.
This is illustrated in Code~\ref{code:branch-before-pcfl} and Code~\ref{code:branch-after-pcfl}.
In can be proven that, with a correct predication algorithm,
the partially linearized code behaves semantically equivalent to the original code.%
\footnote{
  A subset of this transformation was formally proven to be semantics-preserving in CompCert by Knothe et~al.~\cite{knothe2025-pcfl}.
}

\begin{figure}
	\algcaption{Code before linearization}
	\label{code:branch-before-pcfl}
	\begin{algorithmic}
		  \If{pub}
        \If{sec}
          \State{x $\gets$ $y*y$}
        \Else
          \State{x $\gets$ $\sin(y)$}
        \EndIf
		  \EndIf
	\end{algorithmic}
    \abottomrule
	\bigskip

	\algcaption{Code after linearization with $sec$ being secret}
	\label{code:branch-after-pcfl}
	\begin{algorithmic}
		\If{pub}
        \State{$x_1$ $\gets$ $y*y$}
        \State{$x_2$ $\gets$ $\sin(y)$}
        \State{$x$ $\gets$ sec ? $x_1$ : $x_2$}
		  \EndIf
	\end{algorithmic}
    \abottomrule
\end{figure}

In addition, our implementation of \ac{PCFL} can handle arbitrarily nested loops, both with public and with secret exit conditions.
To guarantee constant-time, secret exits are removed and loops are required to exit only through public exits.
Thereby, all exits and affected loop variables are adapted such that the code behaves functionally equivalent.%
\footnote{In rare cases, when a loop exits only through a secret exit and never through a public one before the transformation, it will loop forever after the transformation.
Our pipeline rejects programs with loops that \textit{only} have secret exits.}
Code~\ref{code:example-before-pcfl} and Code~\ref{code:example-after-pcfl} show how
\ac{PCFL} translates a for-loop which can exit on a secret condition.
Note that, if $i$ was reused after the loop (for example to compare it to $n$),
our algorithm would also create a copy of $i$ inside the loop such that its value can be restored
to the value where the loop was originally exited.

\begin{figure}
	\algcaption{Loop before linearization}
	\label{code:example-before-pcfl}
	\begin{algorithmic}
		\For{i=1 {\textbf{to}} n}
		  \If{input[i] $\neq$ password[i]}
		    \State \Return{False}
		  \EndIf
		\EndFor
		\State \Return{True}
	\end{algorithmic}
    \abottomrule
	\bigskip
	\algcaption{After linearization with $password$ being secret}
	\label{code:example-after-pcfl}
	\begin{algorithmic}
		\State{live $\gets$ True}
		\State{result $\gets$ True}
		\For{i=1 {\textbf{to}} n}
		  \State{cond $\gets$ (input[i] $\neq$ password[i])}
		  \State{result $\gets$ live ? result $\wedge$ cond : result}
		  \State{live $\gets$ live $\wedge$ cond}
		\EndFor
		\State \Return{result}
	\end{algorithmic}
    \abottomrule
\end{figure}

When a condition is linearized that originally guarded a memory access, this access is predicated
to ensure the program does not crash after the transformation.
We handle this situation similarly to Soares et~al.~\cite{soares2021lif} by using a dummy memory location
and remark that this may introduce additional cache effects.


\section{Evaluation}
\label{sec:evaluation}

We evaluated our workflow for side-channel elimination with a sample IoT application based on RIOT \cite{baccelli2018RIOT}, a door opener using an AES rolling code.
This application closely mirrors a real-world use case of AES in embedded systems.
We assessed the implementation both empirically using measurements and with static best-case and worst-case timing analysis.
The application receives a packet with an AES-encrypted rolling code via an UART
interface, decodes and verifies the payload, and drives an output according to
the encrypted command.

\subsection{Sample Application}

The AES implementation used in the sample application is, by default, not constant-time.
In particular, subroutine \verb+aes_mul+ (Code~\ref{code:aes_mul}) contains an early out if one of the factors is zero.
Relevant function arguments have been annotated with \verb+maybe_tainted+ attributes using our taint analysis.

The application is run on a small SoC deployed on an FPGA.
The SoC contains a single RISC-V core\footnote{RocketCore configured for
RV32IMCZicsr\_Zifencei\_Zihpm\_Zicond} with an early out for both multiplication
and division. %
Both the data and the instruction cache use the LRU replacement strategy.
Furthermore, the system has 128kB On-Chip SRAM, ensuring deterministic behaviour
regarding instruction loads and memory accesses.

In addition to the external interfaces present in the system---the UART and the pin
driving the actuator---we included a specialized interface for timing measurements.
The interface consists of a memory mapped table of program counter values which may be
set either via the application itself or using a debugger.
Whenever the current PC matches any of the values in that table, an external pin is
pulled high for one clock cycle, allowing for precise timing measurements
by employing software instrumentation and a logic analyzer.


\subsection{Empirical Measurements}


The in-depth statistical analysis is based on Welch's \emph{t}-test and follows the approach of Reparaz et~al.~\cite{dude-ct}.
We collect timing data for one fixed input $I$ and multiple other random inputs $R_i$ that are sent repeatedly via the UART interface to the RIOT application that runs on the RISC-V core.
The timing data for the AES decryption is generated by the application and recorded from the internal debugging console.

The timing data is then evaluated using Welch's \emph{t}-test to determine whether the timing distributions of $I$ and $R_i$ are statistically equivalent.
A t-value larger than $4.5$ provides statistical evidence that the two distributions differ \cite{dude-ct}.
Rather than calculating the t-value once for the entire dataset, we compute it incrementally with each new data point,
allowing us to observe its evolution over time and to identify potential trends.
We repeat this process multiple times using different fixed inputs $I$,
which is important because some fixed inputs may reveal timing leakage more quickly or clearly than others.

\subsection{Static Timing Analysis}

We use \timingprofiler{}, a static timing analysis tool similar to aiT \cite{FerdinandH04}.
In contrast to aiT, which computes safe upper bounds of the \ac{WCET} of safety-critical embedded software and therefore needs a detailed and target-specific model of the hardware characteristics, \timingprofiler{} employs a generic but tunable timing model of a simple processor pipeline including caches.
\timingprofiler{} works as follows:
\begin{enumerate}
    \item First, the fully linked binary executable is disassembled.
          Architecture specific patterns decide whether an instruction is a call, branch, return, or an ordinary instruction.
          Using this knowledge, the control-flow is reconstructed.
          In most cases, this is done completely automatically.
          However, if a target of a call or branch cannot be statically resolved, the user needs to write some annotations to guide the control-flow reconstruction.
    \item Several interprocedural static analyses take place on the \ac{CFG} to determine the values of registers and memory cells, target addresses of memory accesses, and iteration bounds of loops and recursions.
          Moreover, code that can never be executed is marked as infeasible.
          Sometimes, loop bounds cannot be computed statically.
          The user can then guide the analysis via annotations.
    \item Next, a mathematical model of a generic RISC-V processor is used during microarchitectural analysis to derive timing bounds for each instruction.
          This abstract model covers those features that are relevant for timing analysis, incorporating the intrinsic behaviour of a simple 5-stage pipeline including caches and memory.
          \timingprofiler{} allows to adapt its generic timing model, for example by specifying memory access latencies or enabling/disabling caches.
    \item Finally, path analysis searches for the path with either the longest or the shortest execution time.
          Instead of an explicit (and costly) enumeration of all possible pathes, the path analysis constructs an \ac{ILP} using the structural information from the \ac{CFG} and the basic block timing information from the microarchitectural analysis.
          The \ac{ILP} can be generated in such a way that it either minimizes or maximizes the objective function, i.e., the execution time of the function under test.
          The function under test can be considered to be constant-time according to the used timing model in case the best-case and worst-case timing estimates coincide.
\end{enumerate}

\paragraph{Timing Models}
\label{sec:timing_models}

We performed the static timing analysis using six different timing models in the microarchitectural analysis:
\begin{itemize}
    \item \emph{simple} assumes a simple in-order 5-stage pipeline where each instruction takes 1 cycle to execute. There is a single cycle latency for all memory accesses including code fetches.
    \item \emph{memory} is like \emph{simple} but the latency of memory accesses depends on the accessed memory region and supports burst accesses.
    \item \emph{icache} adds 4KB of 4-way set-associative instruction cache with LRU replacement policy to the \emph{memory} timing model, each cache line being 64 bytes wide.
    \item \emph{dcache} adds 2KB of 2-way set-associative data cache with LRU replacement policy to the \emph{memory} timing model, each cache line being 64 bytes wide.
    \item \emph{const-i} ignores pipelining and the memory subsystem. Each instruction takes exactly 1 cycle except for multiplication instructions which take 5 cycles.
    \item \emph{var-i} behaves similar to \emph{const-i}, but multiplication instructions take 2--7 cycles.
\end{itemize}
The four timing models \emph{simple}, \emph{memory}, \emph{icache}, and \emph{dcache} are useful to uncover timing side-channels that are based on timing variances of memory accesses.
However, they ignore any timing variance induced by the execution of an instruction itself (i.e., data-dependent instruction timings).
The timing models \emph{const-i} and \emph{var-i}, on the other hand, ignore timing variances induced by the pipeline and the memory subsystem but focus solely on the instruction execution.
\emph{var-i} allows to take data-dependent instruction timings like \texttt{mul}-early-outs into account.

\section{Results}
\label{sec:results}

\begin{figure}
\centering
\begin{minipage}{\linewidth}
  \centering
  \includegraphics[width=\linewidth]{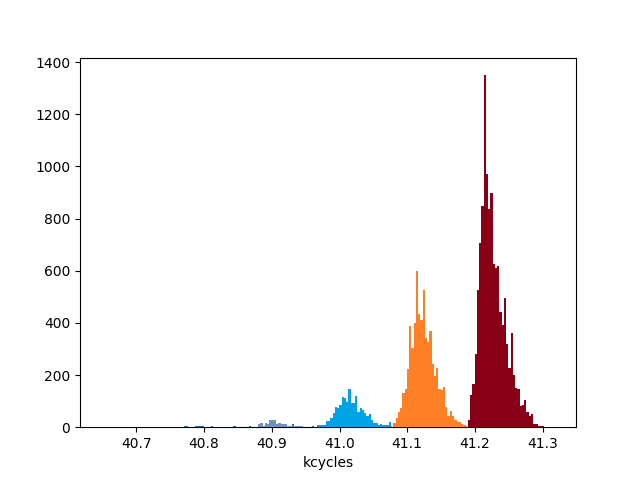}
  \caption{Decryption times without hardening -- the colored spikes approximately follow a Bernoulli distribution with $n=144$ and $p=255/256$}
  \label{results:hist-norm}
\end{minipage} \\

\begin{minipage}{\linewidth}
  \center
  \includegraphics[width=\linewidth]{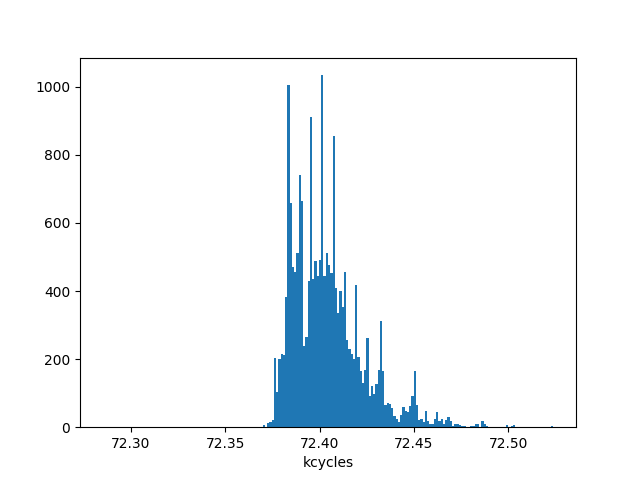}
  \caption{Decryption times with hardening}
  \label{results:hist-ct}
\end{minipage}
\end{figure}

From our sample application, we closely examine the function \verb+aes128_decrypt+.
It calls subroutine \verb+aes_mul+, which is not constant-time as implemented (see Code~\ref{code:aes_mul}). 
\verb+aes_mul+ has two parameters, $a$ and $b$, where only $a$ is marked as tainted.
During decryption of a block, \verb+aes_mul+ is called $144$ times, where the values of $a$ are varying and depend on both the secret key and on the message, while parameter $b$ is public and never zero.
For each call, parameter $a$ seems to approximately follow independent uniform distributions,
so the probability that the early out is taken is $P(a=0) = 1/256$, which yields a Bernoulli distribution of execution times.
This is visible as colored spikes in the histogram compiled from the captured cycle counts
(Figure~\ref{results:hist-norm}) of the baseline, where we compiled the application with CompCert without any constant-time hardening.

\begin{figure}[H]
    \algcaption{Subroutine \emph{aes\_mul}}
    \label{code:aes_mul}
    \begin{lstlisting}[language=C]
// Multiply two numbers in Galois field GF(2^8) 
// with reduction polynomial 0x11B.
__attribute((maybe_tainted(a)))
uint8_t aes_mul(uint8_t a, uint8_t b) {
  if (a && b)
    return Alogtable[(Logtable[a]+Logtable[b])%255];
  else
    return 0;
}
\end{lstlisting}
\end{figure}

After hardening as described in Section~\ref{sec:workflow}, the condition is linearized,
so the exact same sequence of instructions is executed regardless of the value of $a$.
As can be observed in Figure~\ref{results:hist-ct}, while the average execution time
after hardening is higher due to linearization and predication,
the standard deviation drops from $78.8$ to $19.3$ cycles,
and spikes disappear.
Note that our pipeline only linearizes functions that depend on tainted parameters --
in the sample application, \verb+aes_mul+ is the only function that was transformed,
rendering the increase in cycle count less dramatic. 

\paragraph{Static Timing Analysis}

\begin{table}[!th]
    \center
    \begin{tabular}{|l|r|r|}
        \hline
                & baseline                 & hardened \\ \hline\hline
        simple  &  64.1 ..           105.6 &          200.1 \\ \hline
        memory  & 154.4 ..           251.7 &         475.2 \\ \hline
        icache  &  56.9 ..            84.0 & 144.5 .. 148.7 \\ \hline
        dcache  & 142.7 ..           335.5 & 452.0 .. 556.1 \\ \hline
        const-i &  18.9 ..            36.2 &          67.3 \\ \hline
        var-i   &  18.9 ..            38.5 &   63.8 .. 69.6 \\ \hline
    \end{tabular}

    All times in kcycles
    \caption{Static timing analysis results of \texttt{aes128\_decrypt} with and without CFG hardening}
    \label{results:static-both}
\end{table}

Table~\ref{results:static-both} shows the static timing analysis results of \verb+aes128_decrypt+
before and after applying our hardening pipeline.
For the baseline, timing analysis shows differences between best-case and worst-case regardless of the used timing model.
After applying hardening, the best-case and worst-case results coincide for the timing models \emph{simple},
\emph{memory}, and \emph{const-i}, so the code can be considered constant-time in these basic models.
For model \emph{icache} we see some small jitter, which is expected
because the automated hardening does only affect the \ac{CFG} and does not
include any countermeasures against cache effects.
The jitter is however significantly smaller than it was for the baseline.
Without additional information of the CPU the function is executed on,
additional variability on the instruction level may be assumed.
Here, timing model \emph{var-i} uncovers the possible existence of a side-channel due
to the use of multiplication instructions during the computation of the modulo
operation.
Particularly, the multiplication instructions \verb+mul+ and
\verb+mulh+ (which are used for calculating the modulo in Code \ref{code:aes_mul})
can introduce timing variations.

\begin{figure}[!th]
    \center
    \includegraphics[trim={0.7cm 0.1cm 0.5cm 1.5cm},clip=true,scale=0.25]{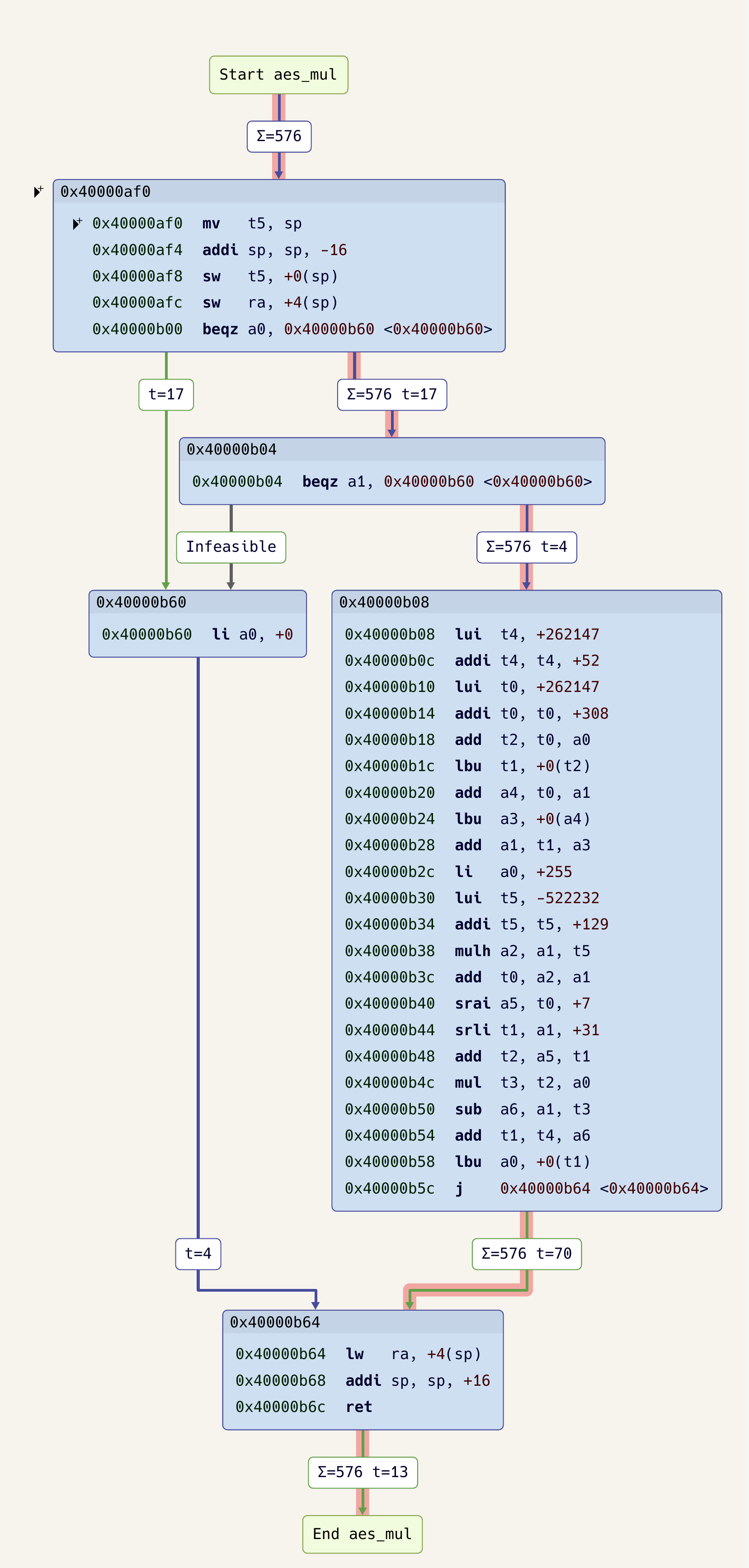}
    \caption{CFG of subroutine \emph{aes\_mul} without hardening}
    \label{fig:aes_mul_original}
\end{figure}

Figure~\ref{fig:aes_mul_original} illustrates the source of execution time
variability in the \ac{CFG} of \verb+aes_mul+.
The worst-case path is highlighted with a red halo.
An early out causes the execution time to be lower if one of the parameters is
zero.

After hardening, the first branch, which depends on the tainted $a$ parameter,
is removed and the rest of the code is predicated.
Therefore, the hardened code executes the same sequence of instructions independently of whether $a=0$ or not.

\paragraph{Statistical Analysis}

\begin{figure}
\centering
\begin{minipage}{\linewidth}
  \centering
  \includegraphics[width=\linewidth]{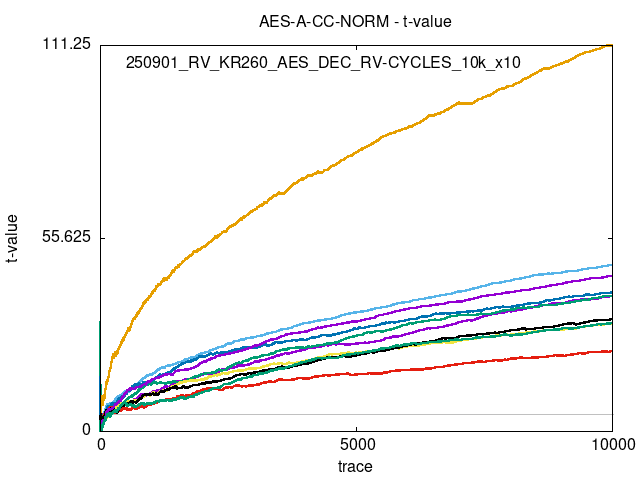}
  \caption{\emph{T}-test results without hardening}
  \label{results:t-test-norm}
\end{minipage} \\
\vspace{1em}
\begin{minipage}{\linewidth}
  \center
  \includegraphics[width=\linewidth]{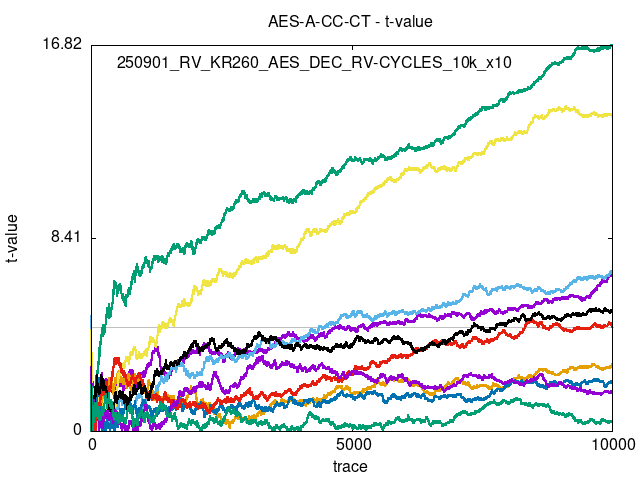}
  \caption{\emph{T}-test results with hardening}
  \label{results:t-test-ct}
\end{minipage} \\
\vspace{1em}
\begin{minipage}{\linewidth}
  \centering
  \includegraphics[width=\linewidth]{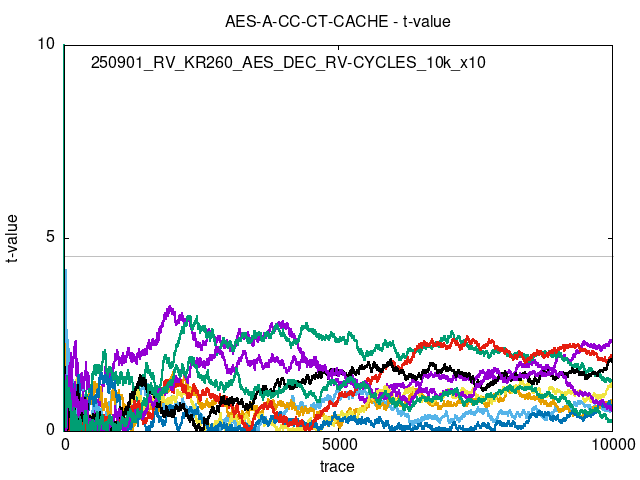}
  \caption{\emph{T}-test results with CFG hardening and countermeasure against cache effects}
  \label{results:t-test-cache}
\end{minipage}
\end{figure}

As expected, the in-depth statistical analysis also indicates a
vulnerability for the baseline.
This can be observed in Figure~\ref{results:t-test-norm}, where the t-value
for every single fixed input (represented by different colors) quickly rises
above the threshold of $4.5$, depicted by a grey line.
After hardening, the t-values are significantly lower, although some do still exceed the threshold
and a general trend for rising values can be observed (Figure~\ref{results:t-test-ct}).
This also suggests the presence of secret-related cache effects that are still present after CFG hardening.

\subsection{Eliminating Cache Effects}

\begin{table}[!ht]
    \center
    \begin{tabular}{|l|r|}
        \hline  & cache-hardened              \\ \hline\hline
        simple  &                     1077.6  \\ \hline
        memory  &                     2545.7 \\ \hline
        icache  &  763.3 ..            770.0 \\ \hline
        dcache  & 2474.0 ..           2767.2 \\ \hline
        const-i &                      371.7 \\ \hline
        var-i   &  354.4 ..             383.2 \\ \hline
    \end{tabular}

    All times in kcycles
    \caption{Static timing analysis results with CFG hardening and countermeasure against cache effects}
    \label{results:static-cache}
\end{table}

These remaining cache effects were eliminated through the following semi-automated source-level change
that was applied before the CFG hardening:
accesses in \verb+aes_mul+ to a single array element in lookup tables (\verb+Logtable+ and \verb+Alogtable+) were
replaced with a loop construct which accesses multiple elements spanning the
entire array in order, such that every cache line is hit exactly once for every array access,
independent of the accessed element.
More precisely, instead of just accessing element \texttt{arr[i]}, we access all elements \texttt{arr[(i mod c) + k*c]},
where $c$ is the length of the cache line, and select the correct value at $k=\lfloor i/c \rfloor$.
We now expect any possibly remaining cache-induced variance
in execution time to be independent from the secret.

As Table~\ref{results:static-cache} indicates, these changes cause a large performance penalty.
The values for \emph{icache} indicate even more pronounced caching effects due
to the increased number of loads and stores, with a spread of 8.4 kcycles versus
only 1.1 kcycles without the additional countermeasures.
This can be explained with the repeated execution of \verb+aes_mul+ as part
of the decryption: one execution of \verb+aes_mul+ leaves the cache in a defined
state with regard to the relevant memory area.
Subsequent executions thus all exhibit the same effects on the cache's state.
However, the static analysis only considers a single, isolated execution of
\verb+aes_mul+ with an unknown initial state of the cache.

The \emph{t}-test, shown in Figure~\ref{results:t-test-cache}, confirms our assessment that any observed variance
is now independent of the secret: no value exceeds the threshold value of 4.5 and no rising trends are identifiable.

\section{Future Work}
\label{sec:future-work}

The aforementioned countermeasure against execution time variance due to cache
effects specifically was carried out semi-manually in source code.
However, for accesses of arrays with a size known at compile time, the process
should be automatable as a compiler transformation,
as the data-flow analysis already allows identification of array
indices that are dependent on secrets.

We did not consider any microarchitectural effects aside from caching effects.
Such effects include individual instructions' execution time
dependence on operands, e.g., an early out commonly implemented in hardware for
multiplication.
These may be addressed during instruction selection within the compiler, given
a specification with execution time for each instruction and, if necessary, patterns
for replacing affected instructions with constant-time software constructs.

\section{Conclusion}
\label{sec:conclusion}

In this paper we demonstrated and evaluated a novel approach for eliminating timing
side-channels from C code with a high degree of automation and compatibility with
current software development flows and build systems.
Effectiveness was shown using static timing analysis and statistical methods such as
Welch's \emph{t}-test.

While the implemented compiler extension eliminated only execution time variance
introduced via control flow, the general approach is also applicable for
handling possible side-channels introduced by the target CPU's microarchitecture
if those effects are known.

\section*{Acknowledgments}

This work was funded by the German Federal Ministry of Research, Technology and Space within the project FreeSBee (funding ID 16KIS1736K).
The responsibility for the content remains with the authors.

\bibliographystyle{plain}
\bibliography{main}

@misc{Pornin2025Pessimist,
  author = {Thomas Pornin},
  title = {{Constant-Time Code: The Pessimist Case}},
  howpublished = {Cryptology {ePrint} Archive, Paper 2025/435},
  year = {2025},
  url = {https://eprint.iacr.org/2025/435},
}

@inproceedings{MollHack2018PCFL,
  author = {Moll, Simon and Hack, Sebastian},
  title = {Partial control-flow linearization},
  booktitle = {Proceedings of the 39th ACM SIGPLAN Conference on Programming Language Design and Implementation},
  pages = {543–556},
  numpages = {14},
  series = {PLDI 2018},
  year = {2018},
  location = {Philadelphia, PA, USA},
  publisher = {Association for Computing Machinery},
  address = {New York, NY, USA},
  doi = {10.1145/3192366.3192413},
  isbn = {9781450356985},
  url = {https://doi.org/10.1145/3192366.3192413},
}

@article{soares2023PCFL,
  author = {Soares, Luigi and Canesche, Michael and Pereira, Fernando Magno Quintão},
  title = {{Side-channel Elimination via Partial Control-flow Linearization}},
  journal = {ACM Trans. Program. Lang. Syst.},
  volume = {45},
  number = {2},
  articleno = {13},
  numpages = {43},
  issue_date = {June 2023},
  month = jun,
  year = {2023},
  publisher = {Association for Computing Machinery},
  address = {New York, NY, USA},
  issn = {0164-0925},
  doi = {10.1145/3594736},
  url = {https://doi.org/10.1145/3594736},
}

@inproceedings{soares2021lif,
  author={Soares, Luigi and Pereira, Fernando Magno Quintãn},
  title={{Memory-Safe Elimination of Side Channels}},
  booktitle={2021 IEEE/ACM International Symposium on Code Generation and Optimization (CGO)},
  pages={200-210},
  year={2021},
  doi={10.1109/CGO51591.2021.9370305},
}

@article{leroy2009compcert,
  author = {Leroy, Xavier},
  title = {Formal verification of a realistic compiler},
  journal = {Commun. ACM},
  pages = {107–115},
  numpages = {9},
  volume = {52},
  number = {7},
  issue_date = {July 2009},
  month = jul,
  year = {2009},
  publisher = {Association for Computing Machinery},
  address = {New York, NY, USA},
  issn = {0001-0782},
  doi = {10.1145/1538788.1538814},
  url = {https://doi.org/10.1145/1538788.1538814},
}

@article{barthe2020compcert-ct-preserving,
  author = {Barthe, Gilles and Blazy, Sandrine and Grégoire, Benjamin and Hutin, Rémi and Laporte, Vincent and Pichardie, David and Trieu, Alix},
  title = {Formal verification of a constant-time preserving {C} compiler},
  journal = {Proc. ACM Program. Lang.},
  articleno = {7},
  numpages = {30},
  volume = {4},
  number = {POPL},
  issue_date = {January 2020},
  month = dec,
  year = {2019},
  publisher = {Association for Computing Machinery},
  address = {New York, NY, USA},
  doi = {10.1145/3371075},
  url = {https://doi.org/10.1145/3371075},
}

@inproceedings{almedia2016ct-verif,
  author = {Almeida, José Bacelar and Barbosa, Manuel and Barthe, Gilles and Dupressoir, Fran\c{c}ois and Emmi, Michael},
  title = {Verifying constant-time implementations},
  booktitle = {Proceedings of the 25th USENIX Conference on Security Symposium},
  series = {SEC'16},
  location = {Austin, TX, USA},
  pages = {53–70},
  numpages = {18},
  year = {2016},
  publisher = {USENIX Association},
  address = {USA},
  isbn = {9781931971324},
}

@inproceedings{borrello2021constantine,
  author = {Borrello, Pietro and D'Elia, Daniele Cono and Querzoni, Leonardo and Giuffrida, Cristiano},
  title = {{Constantine: Automatic Side-Channel Resistance Using Efficient Control and Data Flow Linearization}},
  series = {CCS '21},
  location = {Virtual Event, Republic of Korea},
  pages = {715–733},
  numpages = {19},
  year = {2021},
  publisher = {Association for Computing Machinery},
  address = {New York, NY, USA},
  isbn = {9781450384544},
  doi = {10.1145/3460120.3484583},
  url = {https://doi.org/10.1145/3460120.3484583},
}

@inproceedings{wu2018SC-Eliminator,
  author = {Wu, Meng and Guo, Shengjian and Schaumont, Patrick and Wang, Chao},
  title = {Eliminating timing side-channel leaks using program repair},
  booktitle = {Proceedings of the 27th ACM SIGSOFT International Symposium on Software Testing and Analysis},
  series = {ISSTA 2018},
  location = {Amsterdam, Netherlands},
  pages = {15–26},
  numpages = {12},
  year = {2018},
  publisher = {Association for Computing Machinery},
  address = {New York, NY, USA},
  isbn = {9781450356992},
  doi = {10.1145/3213846.3213851},
  url = {https://doi.org/10.1145/3213846.3213851},
}

@inproceedings{dude-ct,
  author = {Reparaz, Oscar and Balasch, Josep and Verbauwhede, Ingrid},
  title = {Dude, is my code constant time?},
  booktitle = {Proceedings of the Conference on Design, Automation \& Test in Europe},
  series = {DATE '17},
  location = {Lausanne, Switzerland},
  pages = {1701–1706},
  numpages = {6},
  year = {2017},
  publisher = {European Design and Automation Association},
  address = {Leuven, BEL},
}

@inproceedings{kocher1996side-channels,
  author={Paul C. Kocher},
  title={{Timing Attacks on Implementations of Diffie-Hellman, RSA, DSS, and Other Systems}},
  booktitle={Advances in Cryptology - CRYPTO '96, 16th Annual International Cryptology Conference, Santa Barbara, California, USA, August 18-22, 1996, Proceedings},
  series={Lecture Notes in Computer Science},
  volume={1109},
  pages={104-113},
  year=1996,
  publisher={Springer},
  doi={10.1007/3-540-68697-5_9},
}

@article{baccelli2018RIOT,
  author={Baccelli, Emmanuel and Gündoğan, Cenk and Hahm, Oliver and Kietzmann, Peter and Lenders, Martine S. and Petersen, Hauke and Schleiser, Kaspar and Schmidt, Thomas C. and Wählisch, Matthias},
  title={{RIOT: An Open Source Operating System for Low-End Embedded Devices in the IoT}},
  journal={IEEE Internet of Things Journal},
  pages={4428-4440},
  volume={5},
  number={6},
  year={2018},
  doi={10.1109/JIOT.2018.2815038},
}

@InProceedings{knothe2025-pcfl,
  author =	{Knothe, David and Bringmann, Oliver},
  title =	{{On Verifying Secret Control Flow Elimination}},
  booktitle =	{16th International Conference on Interactive Theorem Proving (ITP 2025)},
  pages =	{31:1--31:19},
  series =	{Leibniz International Proceedings in Informatics (LIPIcs)},
  ISBN =	{978-3-95977-396-6},
  ISSN =	{1868-8969},
  year =	{2025},
  volume =	{352},
  editor =	{Forster, Yannick and Keller, Chantal},
  publisher =	{Schloss Dagstuhl -- Leibniz-Zentrum f{\"u}r Informatik},
  address =	{Dagstuhl, Germany},
  URL =		{https://drops.dagstuhl.de/entities/document/10.4230/LIPIcs.ITP.2025.31},
  URN =		{urn:nbn:de:0030-drops-246299},
  doi =		{10.4230/LIPIcs.ITP.2025.31}
}

@inproceedings{BlanchetCCFMMMR03,
  author       = {Bruno Blanchet and
                  Patrick Cousot and
                  Radhia Cousot and
                  J{\'{e}}r{\^{o}}me Feret and
                  Laurent Mauborgne and
                  Antoine Min{\'{e}} and
                  David Monniaux and
                  Xavier Rival},
  editor       = {Ron Cytron and
                  Rajiv Gupta},
  title        = {A static analyzer for large safety-critical software},
  booktitle    = {Proceedings of the {ACM} {SIGPLAN} 2003 Conference on Programming
                  Language Design and Implementation 2003, San Diego, California, USA,
                  June 9-11, 2003},
  pages        = {196--207},
  publisher    = {Association for Computing Machinery},
  year         = {2003},
  url          = {https://doi.org/10.1145/781131.781153},
  doi          = {10.1145/781131.781153},
  bibsource    = {dblp computer science bibliography, https://dblp.org}
}

@inproceedings{KaestnerMWMF22,
TITLE = {{Static Data and Control Coupling Analysis}},
AUTHOR = {K{\"a}stner, Daniel and Mauborgne, Laurent and Wilhelm, Stephan and Mallon, Christoph and Ferdinand, Christian},
URL = {https://hal.science/hal-03694546},
BOOKTITLE = {{11th Embedded Real Time Systems European Congress (ERTS2022)}},
ADDRESS = {Toulouse, France},
YEAR = {2022},
MONTH = Jun,
HAL_ID = {hal-03694546},
HAL_VERSION = {v1},
}

@incollection{FerdinandH04,
author = {Christian Ferdinand and Reinhold Heckmann},
title = {{aiT}: {Worst}-{Case} {Execution} {Time} {Prediction} by {Static} {Programm} {Analysis}},
booktitle = {Building the Information Society. IFIP 18th World Computer Congress, Topical Sessions, 22-27 August 2004, Toulouse, France},
publisher = {Kluwer},
year = {2004},
editor = {Rene Jacquart},
pages = {377-384},
address = {Boston, Mass.},
}

@inproceedings{schneider22,
author = {Schneider, Tobias and Moradi, Amir},
title = {Leakage Assessment Methodology},
year = {2022},
isbn = {978-3-662-48323-7},
publisher = {Springer-Verlag},
address = {Berlin, Heidelberg},
url = {https://doi.org/10.1007/978-3-662-48324-4_25},
doi = {10.1007/978-3-662-48324-4_25},
booktitle = {Cryptographic Hardware and Embedded Systems -- CHES 2015},
pages = {495–513},
numpages = {19}
}

\end{document}